\documentclass[a4paper,11pt,oneside,notitlepage]{article} 

\usepackage{listings}
\usepackage{xcolor}
\usepackage{dingbat}
\usepackage[official]{eurosym}
\usepackage{multirow}
\usepackage{soul}

\usepackage{adjustbox}
 
\definecolor{codegreen}{rgb}{0,0.6,0}
\definecolor{codegray}{rgb}{0.5,0.5,0.5}
\definecolor{codepurple}{rgb}{0.58,0,0.82}
\definecolor{backcolour}{rgb}{0.95,0.95,0.92}
 
\lstdefinestyle{mystyle}{
    backgroundcolor=\color{backcolour},
    commentstyle=\color{codegreen},
    keywordstyle=\color{magenta},
    numberstyle=\tiny\color{codegray},
    stringstyle=\color{codepurple},
    basicstyle=\ttfamily\footnotesize,
    breakatwhitespace=false,         
    breaklines=true,                 
    captionpos=t,                    
    keepspaces=true,                 
    numbers=left,                    
    numbersep=5pt,                  
    showspaces=false,                
    showstringspaces=false,
    showtabs=false,                  
    tabsize=2
}

\usepackage[round]{natbib}

\let\oldbibsection\bibsection
\renewcommand{\bibsection}{\oldbibsection\addcontentsline{toc}{part}{References}}

\usepackage[hyphens]{url}
\usepackage{rotating}
\usepackage{amsmath}
\usepackage{amsthm}
\usepackage{amssymb}
\usepackage{longtable}
\usepackage{tabularx}
\usepackage[hang,bf,singlelinecheck=false]{caption}
\usepackage{float}

\usepackage{endnotes}
\makeatletter
\renewcommand{\@makeenmark}{\textsuperscript{\hyperlink{endnoteAnchor}{[\theenmark]}}}
\makeatother

\usepackage[small,compact]{titlesec}
\titlespacing\section{0pt}{6pt plus 2pt minus 2pt}{0pt plus 1pt minus 0pt}
\titlespacing\subsection{0pt}{6pt plus 2pt minus 2pt}{0pt plus 1pt minus 0pt}
\titlespacing\subsubsection{0pt}{6pt plus 2pt minus 2pt}{0pt plus 1pt minus 0pt}

\usepackage{epstopdf}
\usepackage{paralist}
\usepackage{setspace}
\usepackage[noblocks]{authblk}

\usepackage[colorlinks=true,linkcolor=blue,citecolor=blue]{hyperref}

\usepackage{booktabs,dcolumn} 

\usepackage[left=2.5cm,right=2.5cm,top=3cm,bottom=3cm,includeheadfoot]{geometry} %
\usepackage[flushmargin]{footmisc} %
\usepackage{setspace} %

\makeatletter \@removefromreset{footnote}{chapter}
\makeatother %

\begin{document}


\title{\vspace*{-2em}\Large The risks of risk-based AI regulation: taking liability seriously\thanks{
We are four unrelated academics who work in the same field but from very different disciplinary starting points. Inevitably, we sometimes get confused. We therefore decided to collaborate on this piece and let AI solve the issue of attribution. Authors should be cited in random order. Martin Kretschmer is Professor of Intellectual Property Law and Director of the CREATe Centre, University of Glasgow, UK, martin.kretschmer@glasgow.ac.uk, 10 The Square, Glasgow G12 8QQ, UK. 
Tobias Kretschmer is Professor of Strategy, Technology and Organization, LMU Munich School of Management at Ludwig-Maximilians University of Munich (LMU), Germany, t.kretschmer@lmu.de, Kaulbachstr. 45, D-80539 Munich.
Alexander Peukert is Professor of Civil Law, Commercial Law and Information Law at Goethe University Frankfurt am Main, Germany, a.peukert@jur.uni-frankfurt.de, Max-Horkheimer-Str. 2, Postfach EXC-12, D-60323 Frankfurt am Main.
Christian Peukert is Associate Professor of Digitization, Innovation and Intellectual Property at University of Lausanne (HEC), Lausanne, Switzerland, christian.peukert@unil.ch, Quartier Chamberonne, CH-1015 Lausanne.}}

\author[1]{Martin Kretschmer}
\author[2,3]{Tobias Kretschmer}
\author[4]{Alexander Peukert}
\author[5]{Christian Peukert}

\affil[1]{\small University of Glasgow}
\affil[2]{\small LMU Munich}
\affil[3]{\small CEPR London}
\affil[4]{\small Goethe University Frankfurt am Main}
\affil[5]{\small HEC Lausanne}

\date{\today}

  \maketitle
  \vspace*{-2em}
  \thispagestyle{empty}
\begin{abstract}
\noindent
\onehalfspacing
The development and regulation of multi-purpose, large ``foundation models'' of AI seems to have reached a critical stage, with major investments and new applications announced every other day. Some experts are calling for a moratorium on the training of AI systems more powerful than GPT-4. Legislators globally compete to set the blueprint for a new regulatory regime. This paper analyses the most advanced legal proposal, the European Union’s AI Act currently in the stage of final ``trilogue'' negotiations between the EU institutions. This legislation will likely have extra-territorial implications, sometimes called ``the Brussels effect''. It also constitutes a radical departure from conventional information and communications technology policy by regulating AI ex-ante through a risk-based approach that seeks to prevent certain harmful outcomes based on product safety principles. We offer a review and critique, specifically discussing the AI Act’s problematic obligations regarding data quality and human oversight. Our proposal is to take liability seriously as the key regulatory mechanism. This signals to industry that if a breach of law occurs, firms are required to know in particular what their inputs were and how to retrain the system to remedy the breach. Moreover, we suggest differentiating between endogenous and exogenous sources of potential harm, which can be mitigated by carefully allocating liability between developers and deployers of AI technology. \\

\end{abstract}

\newpage

\thispagestyle{plain}
\setcounter{page}{1} 

\doublespacing

\section{Introduction}

Technology first, then regulation. This has been the received wisdom of Western science and technology policy. Only in exceptional cases, such as nuclear technology or human cloning, do societies ringfence technology development behind ex-ante risk-based obligations that may result in absolute prohibitions in the development of technology.
 
In the digital space, the evolving legal regime since the 1990s fits this logic. Legislators globally decided that liability for the providers of digital technology and services should arise only once there is an actual breach of law (such as ex-post knowledge of illegal material or activity), not through failure to anticipate harm in the first place (ex-ante). For example, imposing publisher-like liability on internet services for user-generated content was considered a barrier to technology development and product and service innovation.\endnote{US Communications Decency Act (CDA) of 1996, Section 230: “No provider or user of an interactive computer service shall be treated as the publisher or speaker of any information provided by another information content provider” (47 USC s. 230).}\endnote{US Digital Millennium Copyright Act of 1998. Section 512 specifies a formal procedure under which service providers are not ex-ante liable but need to respond expeditiously to requests from copyright owners to remove infringing material (notice-and-takedown).}\endnote{EU e-Commerce directive (2000/31/EC) provided a safe harbor for service providers as conduits, caches, and hosts of user information; under Art. 14, ``the service provider is not liable for the information stored at the request of a recipient of the service'' if removed expeditiously upon obtaining relevant knowledge (notice-and-action); Art. 15 prevents the imposition of general monitoring obligations.}
 
However, the consensus around ex-post regulation of digital technology is rapidly dissolving.\endnote{Hacker. P., Engel, A. and Mauer, M. Regulating ChatGPT and Other Large AI Models, \url{https://arxiv.org/abs/2302.02337} (2023).} A global wave of regulation is breaking over developers and deployers of platforms and services. Calls for intervention have intensified since the launch and meteoric rise of ChatGPT in November 2022. Critics have pointed out that ChatGPT’s answers are plausible and confident but often wrong; that users do not know sources of information and how the model was trained; that there is ample potential for unsafe use, including as a tool of misinformation and fraud; and that there are potential infringements of existing rights in personal data and intellectual property.\endnote{It is problematic to locate authoritative citations for the effects of generative AI at this stage. The Economist’s special Science and Technology section (22 April 2023) offers a reasonable overview: \url{https://www.economist.com/interactive/science-and-technology/2023/04/22/large-creative-ai-models-will-transform-how-we-live-and-work}.} In an open letter, scientists and industry actors have called for a 6+ month moratorium for all AI labs on the training of AI systems more powerful than GPT-4 to allow for the development and implementation of a set of shared safety protocols that ensure that systems adhering to them are safe beyond a reasonable doubt.\endnote{Future of Life Institute. Pause Giant AI Experiments: An Open Letter, \url{https://futureoflife.org/open-letter/pause-giant-ai-experiments} (2023).}
 
Indeed, regulators around the globe are currently contemplating such measures. In the United States, responses include a blueprint for an AI Bill of Rights and an AI Risk Management Framework.\endnote{White House Office of Science and Technology Policy, Blueprint for an AI Bill of Rights (2022).}\endnote{National Institute of Standards and Technology, Artificial Intelligence Risk (2023).} In April 2023, the Cyberspace Administration of China circulated draft rules on generative AI services that came into effect on August 15, 2023.\endnote{Cyberspace Administration of China, Interim Measures for the Management of Generative Artificial Intelligence Services (published 13 July 2023), \url{http://www.cac.gov.cn/2023-07/13/c_1690898327029107.htm}.} Yet the most comprehensive and ambitious regulatory activity occurs in the European Union (EU), aiming to shape a new global regime of digital regulation through a stream of far-reaching measures.\endnote{European Commission, A Europe fit for the digital age, \url{https://commission.europa.eu/strategy-and-policy/priorities-2019-2024/europe-fit-digital-age_en} (2020).}
 
We analyze the AI Act proposed by the European Commission and currently in the stages of final trilogue negotiations between the Commission, the Council and the European Parliament through the perspectives of law and economics.\endnote{European Commission, Proposal for a Regulation Laying Down Harmonised Rules on Artificial Intelligence (Artificial Intelligence Act) and Amending Certain Union Legislative Acts, COM/2021/206.}\endnote{Council of the European Union, Proposal for a Regulation of the European Parliament and of the Council laying down harmonised rules on artificial intelligence (Artificial Intelligence Act) and amending certain Union legislative acts - General approach, Council Document 15698/22.}\endnote{European Parliament, Committee on the Internal Market and Consumer Protection and Committee on Civil Liberties, Justice and Home Affairs, Draft Compromise Amendments on the Draft Report Proposal for a Regulation of the European Parliament and of the Council laying down harmonised rules on artificial intelligence (Artificial Intelligence Act) and amending certain Union legislative acts, 16.5.2023.} We argue that the EU approach to extending long-established principles of product safety and cybersecurity to AI via the ‘risk-based approach’ poses specific challenges to a versatile technology. Unlike traditional software, AI systems comprise more than computer code, and, most notably, input data plays a crucial part in determining the output of AI systems. We propose that, instead of working backward from the vast range of potentially harmful outputs of AI to their various root causes, regulation should target and be focused on the inputs, particularly the data used during the operation of AI, which may cause harm to occur. This call for a fundamental change in the regulatory perspective is laid out in three steps. We start with an exposition of the ‘risk-based’ approach to AI pursued by the EU. The next section identifies the limits of this approach, in particular regarding the outsized emphasis on human oversight and the underdeveloped issue of input data management. We conclude by proposing a new ‘liability matrix’ for AI, which distinguishes four corner cases based on the two dimensions of deployment frequency and the ‘locus’ of potential sources of harm arising from inputs such as data and (re)training.
 
\section{The Risk-Based Approach to AI Regulation in the EU}

Though not a global hub of private AI investments\endnote{European Commission, Artificial Intelligence for Europe, COM/2018/237, p. 4.}, the EU is seeking to spearhead the development of global AI norms.\endnote{ European Commission, Fostering a European Approach to Artificial Intelligence, COM/2021/205, p. 4.}\endnote{Bradford A, The Brussels Effect: How the European Union Rules the World (New York, 2020; online edn, Oxford Academic, 19 Dec. 2019), \url{https://doi.org/10.1093/oso/9780190088583.001.0001}.} The European Commission thus proposed an interconnected package of several legal acts concerning AI, extending to hundreds of pages of legislation. This package includes an AI Act laying down horizontal rules on all kinds of AI,\endnote{Legislative procedure 2021/0106/COD.} a revision of sectoral and horizontal product safety rules,\endnote{Regulation (EU) 2023/988 of the European Parliament and of the Council of 10 May 2023 on general product safety, amending Regulation (EU) No 1025/2012 of the European Parliament and of the Council and Directive (EU) 2020/1828 of the European Parliament and the Council, and repealing Directive 2001/95/EC of the European Parliament and of the Council and Council Directive 87/357/EEC, OJ L 135, 23.5.2023, p. 1.} a Cyber Resilience Act addressing cybersecurity issues of AI,\endnote{European Commission, Proposal for a Regulation on horizontal cybersecurity requirements for products with digital elements and amending Regulation (EU) 2019/1020, COM/2022/454 (Cyber Resilience Act), legislative procedure 2022/0272/COD.}, an extension of the Product Liability Directive to software and AI systems\endnote{European Commission, Proposal for a Directive on liability for defective products, COM/2022/495 final, legislative procedure 2022/0302/COD.} and a specific AI Liability Directive.\endnote{European Commission, Proposal for a Directive on adapting non-contractual civil liability rules to artificial intelligence (AI Liability Directive), COM/2022/496 final, legislative procedure 2022/0303/COD.}
 
The starting point of the EU AI regulation is the obvious insight that AI, like every technology, is a double-edged sword. Whereas it can generate a wide array of economic and societal benefits, it can also pose a risk to the safety of persons, their fundamental rights and the general public interest.\endnote{European Commission, Proposal for a Regulation Laying Down Harmonised Rules on Artificial Intelligence (Artificial Intelligence Act) and Amending Certain Union Legislative Acts, COM/2021/206, p. 18.} The purpose of EU AI legislation is to address these risks in ways that help build producer and user trust and social acceptance of these technologies.\endnote{European Parliament, Civil liability regime for artificial intelligence, Official Journal of the European Union, 6.10.2020, C 404/107.} Thus, the EU pursues the twin objectives of protecting fundamental rights and larger societal interests while promoting the uptake of AI to boost the EU’s global competitiveness. The envisaged ‘ecosystem of trust’ is meant to provide companies and public organizations with legal certainty to innovate and implement AI.\endnote{European Commission, White Paper on Artificial Intelligence – A European Approach to Excellence and Trust, COM/2020/65, p. 2.} The Commission presents it as a ‘light governance structure’ that does not disproportionately increase the cost of placing AI solutions on the market.\endnote{European Commission, Fostering a European Approach to Artificial Intelligence, COM/2021/205, p. 6.}
 
To establish an AI ecosystem of trust, EU AI regulation will have a comprehensive scope of application. EU AI legislation will address all AI ‘operators’, including the entities that developed the AI system (‘provider’), professional users, manufacturers of products involving AI, importers and distributors of AI, irrespective of location, as long as they put an AI system on the EU market, or the output produced by the system is used in the Union. Further, it will regulate AI throughout its lifecycle, from design through retirement. EU AI regulation also aims to prevent not only conventional ‘material’ harm to the safety and health of individuals, including loss of life and damage to property but in addition also ‘immaterial’ harm to individual fundamental rights (e.g. loss of privacy, limitations to the right of freedom of expression, human dignity, discrimination for instance in access to employment) and societal interests (e.g. regarding disinformation).\endnote{European Commission, White Paper on Artificial Intelligence – A European Approach to Excellence and Trust, COM/2020/65, p. 10.}
 
The regulatory tools to achieve these goals are not new but are based on concepts of conventional product safety/liability and cybersecurity laws. The idea is to make AI safe and secure like every other hardware or software and to compensate for damages to individuals.\endnote{European Commission, Fostering a European Approach to Artificial Intelligence, COM/2021/205, p. 2.} Consequently, the proposed EU AI legislation contains both ex-ante obligations to ensure safety, cybersecurity, and the protection of fundamental rights as well as ex-post liability rules to compensate damages where an AI risk materializes.\endnote{European Commission, Proposal for a Directive on adapting non-contractual civil liability rules to artificial intelligence (AI Liability Directive), COM/2022/496 final, p. 16.}
 
\begin{figure}[!t]
\caption{The EU’s risk-based approach to AI regulation} 
\label{fig:EU_AI_Pyramid}
\begin{minipage}{\linewidth}
\includegraphics[width=\textwidth,trim=0 330 450 0, clip]{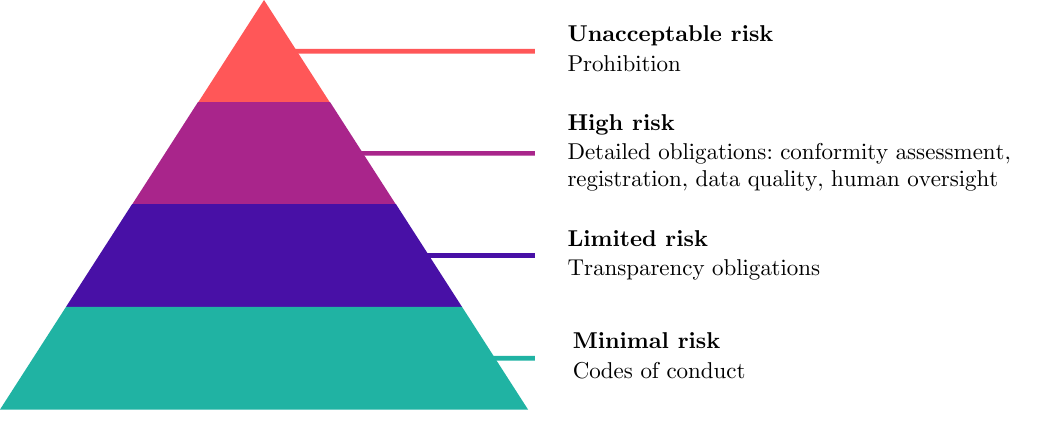}
\end{minipage}\\
\footnotesize{\textbf{Note:} Adapted illustration from the European Commission's regulatory framework proposal on Artificial Intelligence (2022). Source: \url{https://digital-strategy.ec.europa.eu/en/policies/regulatory-framework-ai}.\\ }%
\end{figure}

The former, preventive measures follow a well-established ‘risk-based approach’, which establishes duties of AI operators proportionate to the intensity and scope of the risks that AI systems can generate.\endnote{European Commission, Proposal for a Regulation Laying Down Harmonised Rules on Artificial Intelligence (Artificial Intelligence Act) and Amending Certain Union Legislative Acts, COM/2021/206, p. 9.} In this concept, ‘risk’ functions as the link between an AI system and its potential harm. To account for the characteristic ‘versatility’ of AI\endnote{European Commission, Fostering a European Approach to Artificial Intelligence, COM/2021/205, p. 1.}\endnote{Moor, M., Banerjee, O., Abad, Z.S.H. et al. Foundation models for generalist medical artificial intelligence. Nature 616, 259–265 (2023). \url{https://doi.org/10.1038/s41586-023-05881-4}.} and the wide-ranging scope of potential harms, the EU approach distinguishes between different categories of AI systems and their use (see Figure \ref{fig:EU_AI_Pyramid}).

First, the AI Act singles out certain AI ‘practices’ that will be prohibited upfront. These practices include (1) AI systems deploying subliminal techniques beyond a person’s consciousness to materially distort a person’s behavior, (2) AI systems that exploit vulnerabilities of a specific group of persons due to their age, physical or mental disability, (3) unjustified or disproportionate social scoring systems used by public authorities and (4) the use of ‘real-time’ remote biometric identification systems in publicly accessible spaces for the purpose of law enforcement, unless and in as far as such use is strictly necessary.\endnote{Council of the European Union, Proposal for a Regulation of the European Parliament and of the Council laying down harmonised rules on artificial intelligence (Artificial Intelligence Act) and amending certain Union legislative acts - General approach, Council Document 15698/22, art. 5.}

Second, ‘high-risk’ AI systems are subjected to comprehensive risk management duties, taking into account the generally acknowledged state of the art throughout their entire lifecycle.\endnote{Edwards, L. 2022. The EU AI Act: a summary of its significance and scope, Ada Lovelace Institute, \url{https://www.adalovelaceinstitute.org/ resource/eu-ai-act-explainer}.} Inter alia, AI providers (developers) have to ensure that the high-risk AI system undergoes a conformity assessment procedure and is registered in an EU AI database prior to its placing on the market. There are two categories of high-risk AI. One concerns AI systems that are safety components of products or systems, or which are themselves products or systems subject to existing sectoral safety legislation, for example regarding machinery, medical devices or toys. The second category covers AI systems intended to be used in critical areas, including biometric identification and categorization of natural persons, management and operation of critical infrastructure, education, employment, access to essential private and public services, law enforcement and democratic processes.\endnote{European Commission, Proposal for a Regulation Laying Down Harmonised Rules on Artificial Intelligence (Artificial Intelligence Act) and Amending Certain Union Legislative Acts, COM/2021/206, art. 6(2) and annex III.}
 
Third, EU AI law will establish rules for any other AI system, i.e. any system that uses machine learning and/or logic- and knowledge-based approaches to operate with elements of autonomy, even if that system does not entail an unacceptable or high risk. With regard to these ‘other’ AI systems, the Commission and the Member States are called upon to facilitate codes of conduct in which the signatory providers accept to comply with as many high-risk AI requirements as possible. Moreover, non-high-risk AI systems will be subject to the general product safety rules, which will apply ‘as a safety net’.\endnote{European Commission, Proposal for a Regulation Laying Down Harmonised Rules on Artificial Intelligence (Artificial Intelligence Act) and Amending Certain Union Legislative Acts, COM/2021/206, p. 36.} Consequently, future AI market surveillance authorities may take measures, including a recall of an AI system that poses a risk to the health or safety or to fundamental rights of individuals which goes beyond what is considered reasonable and acceptable in relation to its intended purpose or under the normal or reasonably foreseeable conditions of use of the AI system concerned.\endnote{Council of the European Union, Proposal for a Regulation of the European Parliament and of the Council laying down harmonised rules on artificial intelligence (Artificial Intelligence Act) and amending certain Union legislative acts - General approach, Council Document 15698/22, art. 63.}
 
Finally, the ChatGPT frenzy caused the EU Council (i.e. the EU Member States) and the European Parliament to call for the introduction of even more AI categories. The Council proposes to introduce the category of  ‘general purpose AI’ (GPAI), which is defined as an AI system intended by the provider to perform generally applicable functions such as image and speech recognition, audio and video generation, pattern detection, question answering, translation and others, and thus may be used in a plurality of contexts and be integrated into a plurality of other AI systems.\endnote{Council of the European Union, Proposal for a Regulation of the European Parliament and of the Council laying down harmonised rules on artificial intelligence (Artificial Intelligence Act) and amending certain Union legislative acts - General approach, Council Document 15698/22, p. 62.} GPAI that ‘may’ be used as high-risk AI or as a component of a high-risk AI system shall comply with the requirements established for such high-risk AI irrespective of whether the GPAI is put into service as a pre-trained model and whether further fine-tuning of the model is to be performed by the user of the GPAI. However, GPAI providers can avoid these duties if they have no reason to believe that the system may be misused and explicitly and in good faith exclude all high-risk uses in the instructions of use. Further, the Council proposal empowers the Commission to adapt the high-risk AI requirements applicable to GPAI in light of market and technological developments.\endnote{Council of the European Union, Proposal for a Regulation of the European Parliament and of the Council laying down harmonised rules on artificial intelligence (Artificial Intelligence Act) and amending certain Union legislative acts - General approach, Council Document 15698/22, pp. 70-72.} The European Parliament proposes to introduce yet another category of the subject matter regulated by the AI Act, namely the ’foundation model’. It is defined as a ‘model’ ‘developed from algorithms’ that is trained on broad data at scale, is designed for generality of output, and can be adapted to a wide range of distinctive tasks, performed by specific GPAI systems.\endnote{European Parliament Draft Compromise Amendments on the Draft Report Proposal for a Regulation of the European Parliament and of the Council laying down harmonised rules on artificial intelligence (Artificial Intelligence Act) and amending certain Union legislative acts, 16.5.2023, art. 3(1c), recital 60e.} According to the European Parliament, providers of such foundational models should be subject to specific ex-ante obligations regarding the mitigation of ‘reasonably foreseeable risks’ and the documentation of remaining ‘non-mitigable risks’ after development.\endnote{Art. 28b European Parliament Draft Compromise Amendments on the Draft Report Proposal for a Regulation of the European Parliament and of the Council laying down harmonised rules on artificial intelligence (Artificial Intelligence Act) and amending certain Union legislative acts, 16.5.2023.}
 
As regards the distribution of all these different duties among the AI ‘operators’ involved, the proposals currently on the table again take a differentiated approach, with the overall aim of addressing each obligation to the actor best placed to address any potential risks. Accordingly, risks arising from the development phase must be managed by the providers (including providers who substantially modify an AI system already on the market), whereas separate and targeted duties are set out for importers, distributors and users. The latter will be under a statutory (not only contractual) obligation to use such systems in accordance with the instructions of use accompanying the systems.\endnote{Council of the European Union, Proposal for a Regulation of the European Parliament and of the Council laying down harmonised rules on artificial intelligence (Artificial Intelligence Act) and amending certain Union legislative acts - General approach, Council Document 15698/22, art. 29(1).} The allocation of the duty to compensate damages ex-post follows the same logic.\endnote{European Commission, Proposal for a Directive on adapting non-contractual civil liability rules to artificial intelligence (AI Liability Directive), COM/2022/496 final, p. 18.}
 
\section{Why Regulation Should Focus More on Data and Less on Human Oversight}

The risk-based approach of the proposed AI act creates a tremendous amount of complexity. This complexity is not an unavoidable consequence of the need to regulate AI in a proportionate manner. Instead, it follows from the attempt to extend the traditional product safety approach to AI. The logic of this approach is to work backward from certain harms to measures that mitigate the risk that these harms materialize. Applying this logic to AI is problematic for several reasons:
 
Regarding key inputs, such as data and model training, the recent EU proposal seems underdeveloped and not reflective of some critical technical issues. While the Commission proposal lays out explicit goals for data quality, they seem unrealistic. Actors are required that ``training, validation and testing data sets are relevant, representative, free of errors and complete''.\endnote{European Commission, Proposal for a Regulation Laying Down Harmonised Rules on Artificial Intelligence (Artificial Intelligence Act) and Amending Certain Union Legislative Acts, COM/2021/206, art. 10 para 3.
} These criteria, alone and in combination, are incompatible with core underlying statistical concepts in machine learning.
 
The relevance criterion seems mechanical (without relevant data, in the sense that there are no correlations between input data and desired outputs, no AI system will perform well) and therefore redundant. Completeness needs to be refined further. For example, a dataset could be considered complete if it contains only one feature for the entire population, all features of one subject, or all features of the entire population. In machine learning terminology, the latter would be called ground truth. In the statistical sense, representativeness is related to completeness. A representative dataset reflects the moments (e.g. mean, standard deviation, skewness, etc.) of the distribution of the population, i.e. of a ``complete'' dataset. Finally, whether a dataset is error-free can be determined only if one has a good understanding of what the ground truth is, i.e. with access to either a complete dataset or a representative dataset.
 
Further, we need to distinguish errors in input data that lead to biased model outcomes and those that lead to increased variance in outcomes. Avoiding the former type of error, which reflects an unobserved correlation between error and outcome (e.g., if usage data is only collected at specific times, omitting anyone not online at that time), seems crucial to avoid structural bias in far-reaching applications, e.g., credit scoring.
 
Errors in input data uncorrelated to outcomes (e.g., random lens flare in image data) become critical in applications where inaccuracy is only acceptable within tight margins (e.g., self-driving vehicles), and much less so if AI is used as a complementary system that recommends actions to a human decision maker. Further, applications have varying tolerance for type I and type II errors. Consider the trade-off between freedom of speech and illegal content in content moderation systems. Avoiding type I errors (not removing illegal content) seems preferable to avoiding type II errors (removing lawful content). However, these prediction errors can only be classified if the ground truth is at least approximately known.
 
In addition to the uncertainty around legal terms and their correspondence to well-understood statistical concepts, the proposed AI Act ignores important realities of even the largest input datasets used in current AI systems. Access to data is governed by the decisions of specific actors, making data access subject to strategic behavior. Further, access to data may be enabled or restricted by rules, for example, set by governments. This makes access to data subject to exogenous or regulatory factors outside the control of AI system providers. Even a large language model with access to a gigantic dataset covering all text on the internet only has access to information someone chose to make publicly available. Some information, which might be critical to creating a complete or representative dataset, may be strategically withheld (e.g., journalistic content behind a paywall), not shared because of privacy concerns, or not lawfully usable because of data protection laws (“Right to be forgotten”, GDPR, CCPA, etc.) or intellectual property laws (copyright, trademark, etc.).\endnote{Fiil-Flynn, S.M., Butler, B., Carroll, M., Cohen-Sasson, O., Craig, C., Guibault, L., Jaszi, P., Jütte, B.J., Katz, A., Quintais, J.P. and Margoni, T., 2022. Legal reform to enhance global text and data mining research. Science, 378(6623), pp.951-953.}
 
Absent any relevant data, some models face so-called cold start problems,\endnote{Schein, A.I., Popescul, A., Ungar, L.H. and Pennock, D.M., 2002, August. Methods and metrics for cold-start recommendations. In Proceedings of the 25th annual international ACM SIGIR conference on Research and development in information retrieval (pp. 253-260).} which can only be solved through surrogate data sufficiently correlated with the missing data. Moreover, some of the topics people care about today did not exist yesterday. Perspectives on topics may change as more information becomes available. For instance, science creates new knowledge, which corrects prior misconceptions. The ground truth might change simply because social systems exogenously evolve by nature. As a result, datasets will become stale over time, leading to so-called concept drift, even if they were relevant, representative, free of errors and complete when the AI system was first created.
 
However, data may also be changed strategically by actors, i.e., endogenously, to change the outcome of a model. Examples include tactics to game an algorithm similar to search engine optimization (which one might call AI optimization -- ``AIO'') or adversarial attacks. New business models may allow actors to pay to change the weight of some parameters of a model to achieve a certain outcome, much like advertising payments can affect the ranking in search results on online platforms. Endogenous changes in the ground truth can lead to what machine learning terminology calls data contamination, resulting in biased and lower quality outcomes.
 
In addition to updating datasets, periodically retraining models could offer a potential solution to concept drift and contamination. However, training can be costly, especially for large models. The cost of collecting updated datasets -- in line with regulatory requirements -- and the cost of re-training models create a new economic trade-off for model performance. Poor model performance might affect some users more dramatically than others, and while it might be socially desirable to invest in re-collection and re-training, the cost might exceed the (private) benefits for a unitary actor. These issues resurface in the rules for outcomes and feedback loops laid out in the Commission's AI Act proposal. Article 15 requires that systems need ``an appropriate level of accuracy, robustness and cybersecurity, and perform consistently in those respects throughout their lifecycle'', but it does not reflect key economic trade-offs. A regulation that effectively requires firms to always work off ``perfect’’ data can dissuade investment in AI systems in the first place.
 
Moreover, Article 14 introduces human oversight as a mechanism for risk mitigation. This is highly problematic for several reasons. First, AI systems are often too vast for humans to evaluate every individual outcome. Hence, human oversight would need to be based on representative samples of all outcomes. In some applications, outcomes that must be avoided will be rare, and, therefore, difficult to detect even in a representative sample. Second, even if adverse outcomes were detected by human oversight, in some applications this may be too late to avoid unwanted effects, simply because humans react relatively slowly. Human oversight is therefore only useful for issues that systematically lead to bad outcomes if the systematic nature is detected. Third, human oversight, especially concerning content moderation, can lead to lasting psychological and emotional distress.\endnote{Steiger, M., Bharucha, T.J., Venkatagiri, S., Riedl, M.J. and Lease, M., 2021, May. The psychological well-being of content moderators: the emotional labor of commercial moderation and avenues for improving support. In Proceedings of the 2021 CHI conference on human factors in computing systems (pp. 1-14).} Fourth, recent evidence casts doubt on the fundamental assumption that humans can judge whether algorithms work as intended,\endnote{Jakesch, M., Hancock, J.T. and Naaman, M., 2023. Human heuristics for AI-generated language are flawed. Proceedings of the National Academy of Sciences, 120(11), p.e2208839120.} and many AI systems have been shown to be more accurate than most humans even in complex tasks.\endnote{Katz, D.M., Bommarito, M.J., Gao, S. and Arredondo, P., 2023. Gpt-4 passes the bar exam. Available at SSRN 4389233.} Indeed, human oversight may only perform well enough when carried out by experts with domain knowledge. However, given the scale and scope of AI systems, especially once we approach general-purpose AI systems, there will be substantial demand for human experts to oversee AI systems. Opportunity costs of highly specialized, highly trained humans will be large, especially given that overseeing an algorithm is, at best, not very interesting but likely even detrimental to mental health.
 
\section{Recommendations Based on a New Framework for AI Liability}

Taking a closer look at inputs to AI models as opposed to outputs to define AI risk helps uncover significant differences between AI that gets deployed once and AI that is constantly (re-)deployed and developed. In addition, we believe it is helpful to further differentiate between input-related risks arising from exogenous sources (i.e., independent of the provider’s or the AI users’ actions) and from endogenous sources (e.g., strategic behavior where users or distributors ``game'' the algorithm in a way that results in harmful outcomes).
 
Equally importantly, any legislation on AI has to be flexible enough to accommodate future applications that do not exist yet. If digital transformation has taught us one thing, it is that the creativity of technology providers, complementors, and users has been a driving force in the widespread deployment and use of new digital technologies. It would be naive to expect anything different in AI. A carefully thought-out framework of liability rules that govern responsibility when something goes wrong and defines who needs to act, remedy and possibly pay will be critical to facilitate innovation in AI.\endnote{Hacker, P. The European AI liability directives–Critique of a half-hearted approach and lessons for the future. Computer Law \& Security Review, 51, p.105871. (2023); Buiten, M. Product Liability for Defective AI, https://ssrn.com/abstract=4515202 (2023).}

In this regard too, the EU’ approach is complex. It consists of two separate directive proposals.\endnote{For an extensive analysis see Hacker, P. The European AI liability directives–Critique of a half-hearted approach and lessons for the future. Computer Law \& Security Review, 51, p.105871. (2023).} One directive will set out fully harmonized rules on the strict liability of developers of AI systems for health, property or data damage suffered by natural persons and caused by ``defective'' AI.\endnote{European Commission, Proposal for a Directive on liability for defective products, COM/2022/495 final, legislative procedure 2022/0302/COD.} The other directive will harmonize national fault liability regimes – which will remain applicable next to the strict liability regime – to the extent that users will be granted a right against providers of a high-risk AI system that is suspected of having caused damage to disclose relevant evidence and that a causal link will be presumed to exist between the fault of the provider and the damaging output produced by the AI system.\endnote{European Commission, Proposal for a Directive on adapting non-contractual civil liability rules to artificial intelligence (AI Liability Directive), COM/2022/496 final, legislative procedure 2022/0303/COD.} 

Also as regards these AI liability rules, it seems preferable to us to design sanctions arising from liability for harmful outcomes rather than impose ex-ante restrictions on specific actions. This seems most promising to enforce a responsible use of AI instead of encouraging firms to ``game the system'' and bypass ex-ante rules and regulations.
 
Depending on the context, both the conventional approaches of the product safety model, which allocates all responsibility to the technology developer/provider and the deployer/publisher model in which the deployer of the technology is responsible for risks seem appropriate. Placing too high of a burden on technology developers/providers (vis-a-vis deployers) may (a) give an advantage to larger firms who can shoulder the legal risk more easily, and (b) slow down technological development in ``uncharted territory''. That is, technological developments may be directed at applications for which the developer has low liability, either because most risk is borne by the deployer or because it is considered a low risk application in the first place.
 
\begin{figure}[!t]
\caption{Liability framework focused on AI inputs and deployment models}
\label{tbl:framework}
\begin{tabular}{lll}
\cline{2-3}
\multicolumn{1}{l|}{\begin{minipage}[t]{.4\columnwidth}%
\textbf{Exogenous data issues}\\
\textit{(Cold start, concept drift, privacy policy, etc.)\\}
\end{minipage}}                                         & \multicolumn{1}{l|}{\begin{minipage}[t]{.25\columnwidth}%
Liability of developer

\end{minipage}} & \multicolumn{1}{l|}{\begin{minipage}[t]{.25\columnwidth}%
Liability of developer

\end{minipage}}                    \\ \cline{2-3} 
\multicolumn{1}{l|}{\begin{minipage}[t]{.4\columnwidth}%
\textbf{Endogenous data issues}\\
\textit{(``AIO'', contamination, adversarial attacks, concentration in AI research market, etc.)}
\end{minipage}} & \multicolumn{1}{l|}{\begin{minipage}[t]{.25\columnwidth}%
Liability of deployer
\end{minipage}}  & \multicolumn{1}{l|}{\begin{minipage}[t]{.25\columnwidth}%
Joint liability of developer and deployer
\end{minipage}} \\ \cline{2-3} 
 & \textbf{One-shot deployment} & \textbf{Continuous deployment}\\
\end{tabular}
\end{figure}

In Figure \ref{tbl:framework}, we summarize our thoughts on how liability should be allocated conditional on the characteristics of the AI technology at hand. On the horizontal axis, we think of modes of deployment, mirroring our legal discussion of product safety regulation aspects. At one extreme, we can think of a one-shot deployment, such as a computer vision system to recognize license plate details to manage access of cars to a parking garage. Once the system has learned how to detect license plates in pictures of cars, and how to decipher characters from license plates, there is little need to update underlying datasets because there will be little to no change in how license plates look in the medium to long run. On the other extreme, we can think of continuous deployment, such as a chat system connected to a large language model. For such a system to be truly useful, data needs to be updated relatively frequently, for example, by incorporating user feedback or by adding new data points to the training set.
 
On the vertical axis, we consider data as the key input factor to AI systems, mirroring our technical discussion. We first think of data quality as a function of exogenous issues, such as cold start, concept drift, and privacy regulation. At the other extreme, we can think of data quality as a function of endogenous issues, such as agents trying to “game the system”, for example, with the AI equivalent of search engine optimization (``AIO'') or adversarial attacks that inject biased/selected data. Further, a lack of economic incentives in non-competitive markets for AI research could also lead to endogenous data quality issues if there is not enough investment in the collection and continuous updating of relevant datasets.\endnote{Ahmed, N., Wahed, M. and Thompson, N.C., 2023. The growing influence of industry in AI research. Science, 379(6635), pp.884-886.}

This simple framework allows for a powerful analysis of who should carry the liability or the risk of an AI system. For license plate recognition, the developer should be responsible for issuing an updated system whenever the underlying ground truth changes, for example when a country changes its license plate layout. However, it should be the responsibility of the deployer, the operator of the parking garage in our example, to avoid economic harm from an adversarial attack, such as someone using a fake license plate to enter the parking garage. For a system like Bing AI, if only exogenous issues matter for data quality, again the developer should be liable for investing in minimizing issues such as cold start for certain topics (e.g. human languages underrepresented in standard text corpora such as CommonCrawl or Wikipedia). However, given the feedback loops in continuous deployment from user data, there should be joint liability of the developer and the deployer for minimizing risks from endogenous behavior. For example, both the developer of the underlying large language model and the deployer of the front-facing product should be liable for designing systems that minimize risk from adversarial attacks such as the injection of biased data. 

The ``liability matrix'' in Figure \ref{tbl:framework} illustrates these four polar cases based on the two dimensions of deployment frequency and the ``locus'' of potential sources of harm arising from inputs.
 
Society will want answers on how to correct wrong, infringing or dangerous biases, ``hallucinations'' or ``parroting''.\endnote{Li, Z. Why the European AI Act transparency obligation is insufficient. Nat Mach Intell 5, 559–560 (2023). \url{https://doi.org/10.1038/s42256-023-00672-y}.} AI will be integrated into products, platform services or offered as a service itself, but the last point of service contact with the user may have no knowledge, technological capacity or power to offer a remedy.\endnote{Cobbe, J. and Singh, J. 2021. Artificial Intelligence as a Service: Legal Responsibilities, Liabilities, and Policy Challenges. Computer Law and Security Review, 42. \url{https://doi.org/10.1016/j.clsr.2021.105573}.} Allocating correction liability only to the last point of deployment contact, for example for removing a wrong, unsafe, or libelous return to a search query relying on a large language model, is almost impossible. Even upstream, the model cannot be retrained for each complaint but there will need to be fast (``alignment'') solutions, that may be context-specific and require collaboration between developer and deployer, which a joint liability regime would encourage.

Our liability matrix is also useful for finding an appropriate solution to the unresolved issue of regulating open-source AI.\endnote{See also Borges, G., 2023. Liability for AI Systems Under Current and Future Law: An overview of the key changes envisioned by the proposal of an EU-directive on liability for AI. Computer Law Review International, 24(1), pp.1-8.} Because of its transparency, open-source AI can potentially outperform closed AI systems, especially in terms of robustness and security.\endnote{Empirical evidence on relative security performance of open and closed source software is scarce. The available evidence suggests that open source tends to be quicker in detecting and fixing security issues, but there are clear general patterns, see e.g. Schryen, G., 2009. A comprehensive and comparative analysis of the patching behavior of open source and closed source software vendors. In 2009 Fifth International Conference on IT Security Incident Management and IT Forensics (pp. 153-168) and Schryen, G., 2009. Security of open source and closed source software: An empirical comparison of published vulnerabilities. AMCIS 2009 Proceedings, p.387.} Moreover, examples from other widely used technology categories, such as operating systems, security protocols and web browsers, suggest that open source communities have historically played their part in quickly and continuously fixing vulnerabilities that affected downstream commercial products.\endnote{See for example the analysis of a major vulnerability in OpenSSL in Walden, J., 2020. The impact of a major security event on an open source project: The case of OpenSSL. In Proceedings of the 17th international conference on mining software repositories (pp. 409-419).} That said, using open-sourced AI could still cause harm. Our matrix suggests that in such cases, liability (strict or fault based) should first and foremost lie with the deployer who adapts an open AI for specific purposes and thus triggers endogenous risks.

Society may still consider certain AI systems unacceptable under all circumstances (and therefore prohibited in advance), and find that some AI applications have an unacceptable risk profile, similar to that of nuclear technology. However, the complex ex-ante EU high-risk concept could be replaced with a simple joint obligation for continuously deployed AI to correct unlawful outcomes (under existing laws) expeditiously, and to ensure that this obligation cannot be pushed down by the developer contractually through the value chain to deployers or users.\endnote{Cf. Articles 11 and 13 Proposal for a Directive on liability for defective products, COM/2022/495 final, legislative procedure 2022/0302/COD.} Our liability model suggests how such a simpler solution may be found, supporting the uniquely flexible capabilities of generalist AI.\endnote{Moor, M., Banerjee, O., Abad, Z.S.H. et al. Foundation models for generalist medical artificial intelligence. Nature 616, 259–265 (2023). \url{https://doi.org/10.1038/s41586-023-05881-4}.} Allocating joint liability to developers and deployers for continuously deployed AI systems will provide strong incentives for taking an input- rather than an outcome-based approach to regulating the risks of AI. Only if inputs are sufficiently disclosed should developers be able to escape liability.

\clearpage

\theendnotes
\hypertarget{endnoteAnchor}{}


\end{document}